\documentclass[pre,superscriptaddress]{revtex4}
\usepackage{graphicx,amsmath,amsfonts}
\begin{document}
\title{Some exact results for Boltzmann's annihilation dynamics}
\author{Fran\c{c}ois Coppex}
\affiliation{Department of Physics, University of Gen\`eve, CH-1211 Gen\`eve 4, Switzerland}
\author{Michel Droz}
\affiliation{Department of Physics, University of Gen\`eve, CH-1211 Gen\`eve 4, Switzerland}
\author{Jaros\l aw Piasecki}
\affiliation{Institute of Theoretical Physics, University of Warsaw, PL-00 681 Warsaw, Poland}
\author{Emmanuel Trizac}
\affiliation{Laboratoire de Physique Th\'eorique, B\^atiment 210, Universit\'e de Paris-Sud, 91405 Orsay, France}
\author{Peter Wittwer}
\affiliation{Department of Physics, University of Gen\`eve, CH-1211 Gen\`eve 4, Switzerland}
\pacs{45.05.+x}
\begin {abstract}
The problem of ballistic annihilation for a spatially homogeneous
system is revisited within Boltzmann's kinetic theory in two and three
dimensions.  Analytical results are derived for the time
evolution of the particle density for some isotropic discrete bimodal
velocity modulus distributions.  According to the allowed values of the
velocity modulus, different behaviors are obtained: power law decay
with non-universal exponents depending continuously upon the ratio of
the two velocities, or exponential decay. When one of the two
velocities is equal to zero, the model describes the problem of
ballistic annihilation in presence of static traps. The analytical
predictions are shown to be in agreement with the results of
two-dimensional molecular dynamics simulations.
 
\end{abstract}
\maketitle
%------------------------------------------------------------------------------
\section{introduction}\label{section1}

In ballistically controlled reactions, particles with a given initial
velocity distribution move freely (ballistic motion) in a
$d$-dimensional space.  When two of them meet, they annihilate and
disappear from the system.  This apparently simple problem has
attracted a lot of attention during the past
years~\cite{EF,red1,red2,krap,pia_one,us_one,us_due,rey_one,kafri,evans,sire,
chop, triz,us_three} for the following reasons. First, this is one of
the few problems of nonequilibrium statistical physics that can be
exactly solved in some cases, and second it models some growth and
coarsening processes~\cite{krug_one}.

This field was entered with the pioneering work by Elskens and
Frisch~\cite{EF}, where a one-dimensional system with only two possible
velocities $+c$ or $-c$ was studied.  Using combinatorial analysis,
they showed that the density of particles was decreasing according to a
power law ($t^{-1/2}$) in the case of a symmetric initial velocity
distribution. The investigation of this one-dimensional  problem  was
generalized by Droz~\textit{et al}.~\cite{us_one} to the three-velocity case
where the initial velocity distribution is given by
$\varphi(v;0)=p_+\delta(v-c)+p_0\delta(v)+p_-\delta(v+c)$ with
$p_+=p_-$ (symmetric case) and $p_++p_0+p_-=1$.  It turns out that the
decay of the particle density depends on the details of the initial
velocity distribution.  The following analytical results were obtained.
For $p_0 < 1/4 $, the density $n(v;t)$ of particles with velocity $v=
\{ 0, +c, -c \}$, behaves in the long-time limit as $n(0;t) \sim
t^{-1}, \ n(\pm c;t) \sim t^{-1/2}$.  When  $p_0 = 1 /4 $, $n (0;t)
\sim n(\pm c;t) \sim t^{-2/3}$. Finally, for $p_0 > 1/4$, one finds
that $n(0;t)$ saturates to a nonzero stationary value, while  $n(\pm
c;t)$ decays faster than a power law. Moreover, it was shown that in
one dimension, annihilation dynamics creates strong correlations
between the velocities of colliding particles, which excludes a
Boltzmann-like approximation. Pairs of nearest neighbor particles have
the tendency to align their velocities and propagate in the same
direction~\cite{us_one}.

An analytical investigation of the one-dimensional case with a continuous
velocity distribution is much more difficult. A dynamical scaling
theory, whose validity was supported by extensive numerical simulations
for several velocity distributions, led Rey~\textit{et al}.~\cite{us_due} to the
conjecture that all the continuous velocity distributions $\varphi(v)$
that are symmetric, regular and, such that $\varphi(0) \neq 0$ are
attracted in the long-time regime towards the same Gaussian-like
distribution and thus belong to the same universality class.

For higher dimensions, most of the studies are based on an uncontrolled
Boltzmann-like description~\cite{red1,red2,sire} or numerical
simulations~\cite{chop}.  However, based on phenomenological mean-field-like arguments,  Krapivsky \textit{et al}. have studied the
annihilation kinematics of a bimodal velocity modulus distribution in
$d> 2$ dimensions~\cite{krap}. In the case of a mixture of moving and
motionless particles they showed that the stationary particles always
persist, while the density of moving particles decays exponentially.
This approach contains unknown phenomenological parameters, and thus a complete
comparison with the results obtained by numerical simulation is not
possible.

In a recent paper, Piasecki~\textit{et al}.~\cite{us_three} gave an analytical derivation of the hierarchy equations obeyed by the reduced distributions for the annihilation dynamics. In dimension $d >1$ for a spatially homogeneous system, and in the limit (the so-called Grad limit) for which the particle diameter $\sigma \to 0$, and the particle density $n(t) \to \infty$ such that $n(t)\sigma^{d-1}=\lambda^{-1}$, where $\lambda$ is the mean free path, the hierarchy reduces to the Boltzmann-like hierarchy. This hierarchy propagates the factorization of the reduced $k$-particle distribution in terms of one-particle distribution functions. Thus, if  the initial state is factorized, the whole hierarchy reduces to one nonlinear equation for the one-particle Boltzmann distribution.
For annihilation kinetics, the ratio of particle diameter to mean free path vanishes in the long-time limit and the situation becomes similar to the Grad limit discussed above for $\lambda \to \infty$. Thus the long-time limit of the annihilation dynamics (for $d>1$) is likely to be adequately described by the nonlinear Boltzmann equation.

A scaling analysis of the nonlinear Boltzmann equation led to analytical expressions for the exponents describing the decay of the particle density and of the root-mean-square velocity in the case of continuous velocity distributions~\cite{us_three}.

In view of the different behaviors observed in one dimension for
discrete or continuous velocity distributions, it is relevant to study
the case of distributions with discrete modulus spectrum in dimensions
higher than $1$. The goal of this paper is to investigate simple
examples of this kind in three dimensions for which the non linear
Boltzmann equation derived in Ref.~\cite{us_three} can be exactly solved.
The generalization of this approach to an arbitrary dimension is
straightforward, and for the sake of comparison with numerical
simulations, we shall also consider the two-dimensional situation in
some detail.

The validity of the Boltzmann description in the long-time limit
will be confirmed by comparing our analytical predictions with the results obtained by a molecular dynamics simulation.

The paper is organized as follows. In Sect.~\ref{section2} we define the model. 
In Sect.~\ref{section3} the three-dimensional Boltzmann equation is solved 
analytically for a two velocity modulus ($c_1$ and $c_2$) isotropic 
distribution. For simplicity we first consider the one velocity model $c_1=c_2 > 0$ that 
allows to draw interesting comparisons with the same model in one dimension. 
Then the implicit solution for the particle densities in the general case 
$c_1 > c_2 > 0$ is established. It is shown analytically that in the long-time limit the particle densities decay according to power laws, with exponents 
depending continuously on the value of the velocity modulus ratio. We also find 
upper and lower bounds to the particle densities that are compared with the 
numerical solution of the dynamical equation. The particular case of a mixture 
of moving ($c_2>0$) and motionless ($c_1 = 0$) particles is also investigated. 
It turns out that the particle densities decay exponentially to zero for the 
moving particles, and to a nonzero value for the motionless ones. 
This phenomenology is independent of space dimension, and in Sect.~\ref{section:md},
it will be shown explicitly to hold in two dimensions by implementing 
molecular dynamics simulations. This numerical method has the advantage of
being free of the approximations underlying Boltzmann's dynamics and, therefore,
provides an interesting test for the analytical predictions.
Section~\ref{section4} contains our interpretations and conclusions.

%------------------------------------------------------------------------------
\section{The model}\label{section2}

We consider a system made of spheres of diameter $\sigma$ moving ballistically in three-dimensional space. If two particles touch each other, they annihilate and thus disappear from the system. We consider only two-body collisions. The initial spatial distribution of particles is supposed to be uniform, therefore it remains uniform during the evolution. Indeed, if the state is initially translationally invariant, then the free evolution preserves this symmetry. Annihilation dynamics adds the effect of binary collisions that depends only on the distance between particles, thus preserving the translational invariance. Finally, existing numerical simulations seem to be compatible with this assumption of homogeneity~\cite{us_three}. We are interested in the time evolution of the number density of particles with a given velocity modulus.

Let $f_1(\mathbf{v};t)$ be the distribution function of the density of particles in $\mathbb{R}^3$ with velocity $\mathbf{v} \in \mathbb{R}^3$ at time $t$. For spatially homogeneous states, the distribution function has the form
\begin{equation}
f_1(\mathbf{v};t) = n(t)\varphi(\mathbf{v};t), \label{f1}
\end{equation}
where $\varphi(\mathbf{v};t)$ is the velocity probability density. In the long-time limit, Piasecki~\textit{et al}.~\cite{us_three} have shown that the hierarchy satisfied by the reduced distributions approached the Boltzmann hierarchy.
If the initial state is factorized, the nonlinear Boltzmann equation provides then the complete description of annihilation dynamics,
\begin{equation}
\frac{\partial}{\partial t} f_1(\mathbf{v}_1;t) =  \sigma^{2} \int d\widehat{\boldsymbol{\sigma}} \, \, \theta(-\widehat{\boldsymbol{\sigma}} \cdot \widehat{\mathbf{v}}_{12}) (\widehat{\boldsymbol{\sigma}} \cdot \widehat{\mathbf{v}}_{12}) \int_{\mathbb{R}^3} d\mathbf{v}_2 \, |\mathbf{v}_{12}| f_1(\mathbf{v}_1;t) f_1(\mathbf{v}_2;t). \label{boltz1}
\end{equation}
Here $\theta$ is the Heaviside function, 
$\mathbf{v}_{12} = \mathbf{v}_1-\mathbf{v}_2$ the relative velocity of two particles, 
$\widehat{\mathbf{v}}_{12} = \mathbf{v}_{12}/v_{12}$ a unit vector, 
$v_{12}=|\mathbf{v}_{12}|$, and the integration with respect to 
$d\widehat{\boldsymbol{\sigma}}$ is the angular integration over the solid 
angle.

We consider spherically symmetric initial conditions $f_1(v;0)$, $v = |\mathbf{v}|$. This symmetry property is propagated by the dynamics. The Boltzmann equation (\ref{boltz1}) then takes the form
\begin{equation}
\frac{\partial}{\partial t} f_1(v;t) = - \tfrac{2}{3} (\pi \sigma)^2 f_1(v;t) \int_0^\infty du \, u^{2} \, f_1(u;t) \left[ \frac{(u+v)^3-|u-v|^3}{u \, v} \right]. \label{boltz2}
\end{equation}
Equation~(\ref{boltz2}) is a nonlinear homogeneous integral equation for the 
distribution function $f_1(v;t)$. A simplification arises if the initial velocity 
distribution has a discrete modulus spectrum. This spectrum is preserved by the 
annihilation dynamics as no new velocities are created. A simple case is 
provided by the bimodal distribution
\begin{equation}
\varphi(v,0) = \frac{A}{4 \pi c_1^{2}} \delta(v-c_1) + \frac{1-A}{4 \pi c_2^{2}} \delta(v-c_2), \label{initial}
\end{equation}
where $c_2 > c_1 \geq 0$ and $A$ denotes the fraction of particles with velocity modulus $c_1$.

%------------------------------------------------------------------------------
\section{Exact results}\label{section3}
Before addressing the general case, we first consider the single-species problem where $c_2 = c_1 > 0$.

\subsection{Single-velocity modulus distribution}\label{section3-1}
%-----------------------------------------------
Setting $c_2 = c_1 = c > 0$ in Eq.~(\ref{initial}), one obtains from Eq.~(\ref{f1})
\begin{equation}
f_1(v;t) = n(t) \frac{1}{4\pi c^{2}} \delta(v-c). \label{3-1-1}
\end{equation}
From the kinetic equation~(\ref{boltz2}), we find
\begin{equation}
\frac{d}{dt}n(t) = - \tfrac{4}{3} \pi \sigma^{2} c \, n(t)^2, \label{3-1-2}
\end{equation}
whose solution is
\begin{equation}
n(t) = \frac{n_0}{1+\tfrac{4}{3} \pi\sigma^{2} n_0 c t}, \label{3-1-3}
\end{equation}
where $n(0) = n_0$.
A striking observation is that, in the limit $t \to \infty$, the 
density~(\ref{3-1-3}) becomes independent of its initial value $n_0$. 
Note that the same 
phenomenon is also present for simple diffusion limited annihilation such as 
$A + A \to 0$, when the dimension of the system is larger than 
$2$~\cite{reactiondiffusion}.

Contrary to the one-dimensional case for which it has been rigorously shown that the density decays 
proportionally to $t^{-1/2}$~\cite{EF}, one sees from Eq.~(\ref{3-1-3}) 
that in three dimensions, Boltzmann's dynamics is faster as the density 
decays according to $t^{-1}$, which is the mean-field value~\cite{krap}.
We note, however, that the same behavior
$n(t) \propto 1/t$ holds in all dimensions within Boltzmann's kinetic theory 
(and in fact, more generally within the framework of a scaling analysis of 
the hierarchy governing the dynamics of ballistic annihilation~\cite{us_three}).
This discrepancy between Boltmann's prediction and the exact result 
in one dimension illustrates the crucial importance of dynamical correlations
when $d=1$. On the other hand, as suggested in Ref.~\cite{us_three} and 
explicitly shown below by molecular dynamics simulations, the
nonlinear Boltzmann equation is relevant for describing the long-time
dynamics of ballistic annihilation when $d\geq 2$. In this case the particles are very diluted and no dynamical correlations can develop during the time evolution, which would violate the molecular chaos hypothesis.

\subsection{Mixture of particles with two nonzero velocity moduli}\label{section3-2}
%-----------------------------------------------
Consider the case where particles with velocities $c_1 > 0$ and $c_2 > c_1$ are initially present. Thus $f_1(v;t)$ is of the form
\begin{equation}
f_1(v;t) = X(t) \frac{1}{4\pi c_1^{2}} \delta(v-c_1) + Y(t)\frac{1}{4\pi c_2^{2}} \delta(v-c_2), \label{3-2-1}
\end{equation}
where $X(t)$ and $Y(t)$ are, respectively, the densities of particles with velocities $c_1$ and $c_2$. They add up to the total density $X(t) + Y(t) = n(t)$. Upon rescaling the time according to $\tau = t c_2 \pi \sigma^2 / 3$, it follows from Eq.~(\ref{boltz2}) that
\begin{subequations}
\label{allsyst1}
\begin{eqnarray}
\dot{X}(\tau) &=& - 4 \gamma X(\tau)^2 - (3+\gamma^2) X(\tau)Y(\tau), \label{3-2-2a} \\
\dot{Y}(\tau) &=& - 4 Y(\tau)^2 - (3+\gamma^2) X(\tau)Y(\tau), \label{3-2-2b}
\end{eqnarray}
\end{subequations}
where $0 \leq \gamma = c_1/c_2 < 1$, and the overdot denotes time derivative with respect to $\tau$. 

The set of equations~(\ref{allsyst1}) is a nonlinear homogeneous system of coupled differential equations with constant coefficients. An implicit solution can be obtained by introducing the function $V(\tau)$ defined as $V(\tau)=Y(\tau)/X(\tau)$. From Eq.~(\ref{allsyst1}) we get
\begin{equation}
\frac{d Y}{d X} = \frac{4 V^2 +(3+\gamma^2)V}{4 \gamma + (3+\gamma^2)V},
\end{equation}
so that
\begin{equation}
\frac{dX}{X} = \frac{4 \gamma + (3+\gamma^2)V}{(1-\gamma^2)V^2+(1-\gamma)(3-\gamma)V} \, dV. \label{3-2-2c}
\end{equation}
Integrating Eq.~(\ref{3-2-2c}) yields
\begin{equation}
\frac{X_0}{X} = \left( \frac{V_0}{V} \right)^\alpha \left( \frac{V_0 + \tfrac{3-\gamma}{1+\gamma}}{V + \tfrac{3-\gamma}{1+\gamma}} \right)^\beta, \label{3-2-3}
\end{equation}
with $\alpha = 4 \gamma/[(1-\gamma)(3-\gamma)] \geq 0$, $\beta = (3+\gamma^2)/(1-\gamma^2) - \alpha > 0$, $V(0) = V_0=Y_0/X_0$, $X_0 = X(0)$, $Y_0 = Y(0)$. The special case of $\gamma = 0$ will be discussed in Sec.~\ref{section3-3}, hence from now on we assume that $\gamma > 0$, so that $\alpha > 0$. Equations~(\ref{allsyst1}) can also be written as
\begin{subequations}
\label{allsyst2}
\begin{eqnarray}
\frac{d}{d \tau} \left(\frac{1}{X}\right) &=& 4 \gamma + (3+\gamma^2) \frac{Y(\tau)}{X(\tau)}, \label{3-2-4a} \\
\frac{d}{d \tau} \left(\frac{1}{Y}\right) &=& 4 + (3+\gamma^2) \frac{X(\tau)}{Y(\tau)}.\label{3-2-4b}
\end{eqnarray}
\end{subequations}
Multiplying the right-hand side (RHS) of Eq.~(\ref{3-2-4a}) by $X_0$ and equating it with the derivative of the RHS of Eq.~(\ref{3-2-3}), one obtains upon integration from $0$ to $\tau$ the relation
\begin{equation}
X_0 \tau = \int_{V}^{V_0} d u \, \frac{1}{4\gamma + (3+\gamma^2) u} \left\{ -\frac{d}{d u} \left[ \left( \frac{V_0}{u} \right)^\alpha \left( \frac{V_0 + \tfrac{3-\gamma}{1+\gamma}}{u + \tfrac{3-\gamma}{1+\gamma}} \right)^\beta \right] \right\}. \label{3-2-5}
\end{equation}
Equation~(\ref{3-2-5}) implicitly defines the time dependence of the function $V(\tau)$. The procedure to obtain the densities $X(\tau)$ and $Y(\tau)$ from Eq.~(\ref{3-2-5}) is as follows. The integration in Eq.~(\ref{3-2-5}) leads to Appel functions, that may be inverted (at least numerically) in order to give $V(\tau)$. The insertion of $V(\tau)$ in Eq.~(\ref{3-2-3}) 
then gives $X(\tau)$. It is then straightforward to obtain $Y(\tau)$, having determined $V(\tau)$ and $X(\tau)$. The structure of the implicit relation~(\ref{3-2-5}) permits us to establish interesting analytical results.

First, let us investigate the long time behavior of the particle densities $X(\tau)$ and $Y(\tau)$. When $\tau \to \infty$, the LHS of Eq.~(\ref{3-2-5}) diverges linearly which implies that $\lim_{\tau \to \infty} V(\tau) = 0$. So, in the long-time limit, the implicit relation~(\ref{3-2-5}) leads to the asymptotic formula
\begin{equation}
X_0 \tau \stackrel{\tau \to \infty}{\simeq} \frac{1}{4 \gamma} \left( \frac{V_0}{V} \right)^\alpha \left( \frac{1+\gamma}{3-\gamma} V_0 + 1 \right)^\beta, \label{3-2-5b}
\end{equation}
so that
\begin{equation}
V(\tau) \stackrel{\tau \to \infty}{\simeq} V_0 \left( \frac{1+\gamma}{3-\gamma}V_0 + 1 \right)^{\beta/\alpha} \left( 4 \gamma X_0 \tau \right)^{-1/\alpha}. \label{3-2-5c}
\end{equation}
As $\lim_{\tau \to 0} V(\tau) = 0$, Eq.~(\ref{3-2-4a}) takes the asymptotic form $d/d\tau(1/X) = 4 \gamma$. Hence we conclude that
\begin{equation}
X(\tau) \stackrel{\tau \to \infty}{\simeq} \tfrac{1}{4 \gamma} \tau^{-1}. \label{3-2-5e}
\end{equation}
On the other hand, from Eqs.~(\ref{3-2-4b}) and~(\ref{3-2-5c}) we find the long-time relation
\begin{equation}
\frac{d}{d \tau} \left(\frac{1}{Y}\right) \stackrel{\tau \to \infty}{\simeq} (3+\gamma^2) \frac{1}{V_0} \left( \frac{1+\gamma}{3-\gamma}V_0 + 1 \right)^{-\beta/\alpha} \left( 4 \gamma X_0 \tau \right)^{1/\alpha}, \label{3-2-5f}
\end{equation}
which upon integration yields
\begin{equation}
Y(\tau) \stackrel{\tau \to \infty}{\simeq} \frac{V_0}{4 \gamma} (4 \gamma X_0)^{-1/\alpha} \left( \frac{1+\gamma}{3-\gamma}V_0 + 1 \right)^{\beta/\alpha}\tau^{-(3+\gamma^2)/4 \gamma}. \label{3-2-5g}
\end{equation}
Note that the exponent for the density $Y(\tau)$ is a function of the ratio $\gamma = c_1/c_2$ and thus is nonuniversal. In the limit $\gamma \to 1$ one recovers the asymptotic behavior of the single-velocity modulus distribution (see Sect.~\ref{section3-1}).

\begin{figure}[ht]
\begin{center}
\includegraphics[width=0.6\columnwidth]{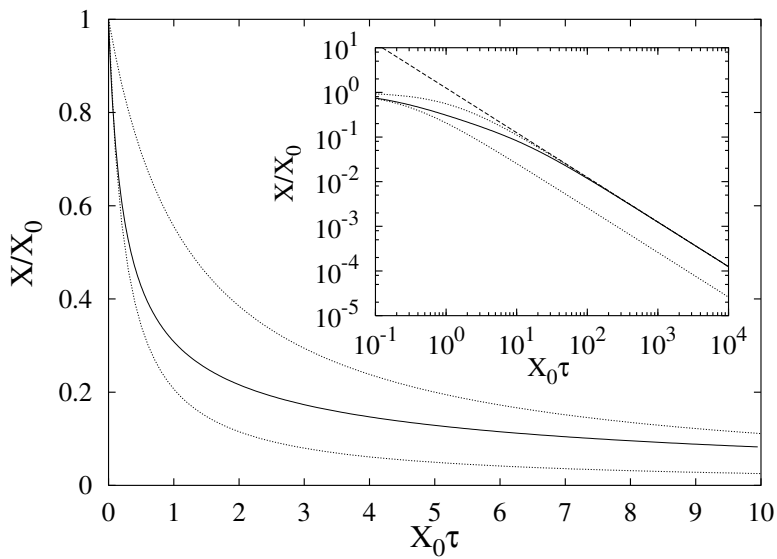}
\end{center}
\caption{\label{fig1} Upper and lower bounds (\ref{3-2-6}) (dotted lines) as well as the numerical solution of the set of equations~(\ref{allsyst1}) for $X(\tau)$ with $X_0 = Y_0$, $\gamma = 0.2$ (continuous line). The inner logarithmic plot shows indeed the power law behavior $X(\tau) \sim \tau^{-1}$ for $\tau \to \infty$, where the asymptotic solution~(\ref{3-2-5e}) is represented by the dashed straight line. Moreover, in this regime the solution converges to the upper bound~(\ref{3-2-6}).}
\end{figure}

\begin{figure}[ht]
\begin{center}
\includegraphics[width=0.6\columnwidth]{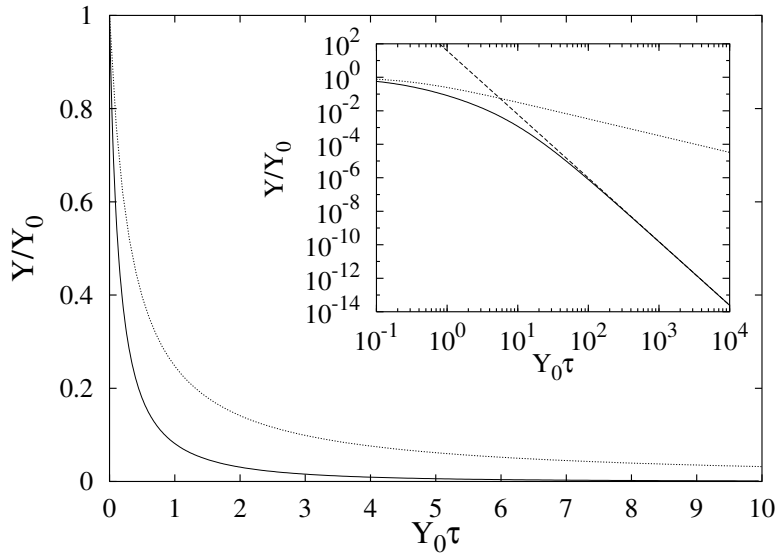}
\end{center}
\caption{\label{fig2} Upper bound (\ref{3-2-8}) (dashed lines) as well as the numerical solution of the set of equations~(\ref{allsyst1}) for $Y(\tau)$ with $X_0 = Y_0$, $\gamma = 0.2$ (continuous line). The inner logarithmic plot of the numerical solution shows indeed the power law behavior $Y(\tau) \sim \tau^{-(3+\gamma^2)/4 \gamma}$ for $\tau \to \infty$, where the asymptotic solution~(\ref{3-2-5g}) is represented by the dashed straight. Furthermore, the use of both the upper bound~(\ref{3-2-5e}) and the asymptotic form~(\ref{3-2-5g}) allows to find an analytical approximation for $Y(\tau)$, which turns out to be exact in both limits $\tau \to 0$ and $\tau \to \infty$.}
\end{figure}

Second, we may find analytical upper and lower bounds for $X(\tau)$ and $Y(\tau)$. Granted that $\alpha > 0$ and $\beta > 0$, the integrand of Eq.~(\ref{3-2-5}) is a strictly monotonic decreasing positive function of $u$, therefore $V(\tau) < V_0$ for all $\tau > 0$. Considering that $(4 \gamma)^{-1} \geq [4\gamma + (3+\gamma^2)u]^{-1}$ for $u \geq 0$, the insertion of Eq.~(\ref{3-2-3}) in Eq.~(\ref{3-2-5}) provides the inequality $X_0 \tau \leq \left( X_0/X - 1 \right)/4\gamma$, which leads to an upper bound for $X(\tau)$. On the other hand, the inequality $(4\gamma+(3+\gamma^2)u)^{-1} \leq [4\gamma+(3+\gamma^2)V_0]^{-1}$ yields a lower bound, so that we finally get
\begin{equation}
\frac{X_0}{1+[4\gamma X_0 + (3+\gamma^2)Y_0]\tau} \leq X(\tau) \leq 
\frac{X_0}{1+4\gamma X_0 \tau}. \label{3-2-6}
\end{equation}
Note that for times such that
\begin{equation}
4 \gamma X_0 \tau \gg 1, \label{caractime}
\end{equation}
the upper bound~(\ref{3-2-6}) coincides with the exact asymptotic 
relation~(\ref{3-2-5e}).
The same kind of analysis as that leading to Eq.~(\ref{3-2-6}) yields the upper bound,
\begin{equation}
0 \leq Y(\tau) \leq \frac{Y_0}{1+(3+\gamma^2)X_0 \tau}. \label{3-2-8}
\end{equation}
The width defined by the difference of the bounds in both cases (\ref{3-2-6}) and (\ref{3-2-8}) is  $\mathcal{O}\left( \tau^{-1} \right)$. Figures~\ref{fig1} and~\ref{fig2} show the numerical solution for $X(\tau)$, $Y(\tau)$, the bounds~(\ref{3-2-6}) and~(\ref{3-2-8}), as well as their asymptotic behaviors~(\ref{3-2-5e}) and~(\ref{3-2-5g}) on a logarithmic scale.

The knowledge of the numerical solution (see Figs.~\ref{fig1} and~\ref{fig2}) allows to determine the crossover time, separating the early and long-time (power law) regimes.

\subsection{Mixture of moving and motionless particles}\label{section3-3}
%-----------------------------------------------
We now consider a particular case of Sec.~\ref{section3-2} that we solve exactly in the asymptotic limit $\tau \to \infty$. The system is now characterized by a certain number of motionless particles (zero velocity, $c_1 = 0$) whereas the rest of the particles have a given nonzero velocity modulus. Thus, setting $\gamma = 0$ in Eq.~(\ref{3-2-3}), inverting the relation in order to find $V=V(X)$, then making use of Eq.~(\ref{3-2-4a}) with $\gamma = 0$ leads to
\begin{equation}
\frac{d}{d \tau} \left( \frac{1}{X} \right) = 3 \left( 3 + \frac{Y_0}{X_0} \right) \left( \frac{X}{X_0} \right)^{1/3} - 9. \label{3-3-1}
\end{equation}
The integration of Eq.~(\ref{3-3-1}) yields
\begin{equation}
3 X_0 \tau = \tfrac{1}{3} \left[1-\tfrac{X_0}{X} \right] + \tfrac{a}{2} \left[ 1- \left( \tfrac{X_0}{X} \right)^{2/3} \right] + a^2 \left[ 1- \left(\tfrac{X_0}{X}\right)^{1/3} \right] + a^3 \ln\left[ \tfrac{a-1}{a-\left( X_0/X \right)^{1/3}} \right], \label{3-3-2}
\end{equation}
with $a=1+Y_0/3 X_0$. In the asymptotic limit $\tau \to \infty$, the LHS of Eq.~(\ref{3-3-2}) tends to $+\infty$. The density $X(\tau)$ cannot thus tend to zero, and must approach a strictly positive value $X(\infty) = X_\infty > 0$. For $\tau \to \infty$, all terms on the RHS of Eq.~(\ref{3-3-2}) but the logarithmic one approach a finite limit. This implies the asymptotic behavior
\begin{equation}
X_\infty = \lim_{\tau \to \infty} X(\tau) = \frac{X_0}{a^3} = X_0 \left( \frac{3 X_0}{3 X_0 + Y_0} \right)^3 > 0. \label{3-3-3}
\end{equation}
Replacing $X$ by $X_\infty$ in all terms of Eq.~(\ref{3-3-2}) except the logarithmic one and then inverting the relation $\tau(X)$, we find
\begin{equation}
X(\tau) \, \stackrel{\tau \to \infty}{\simeq} \, X_\infty \left[  1-\varepsilon(X_0,Y_0;\tau)  \right]^{-3} \, \stackrel{\tau \to \infty}{\simeq} \, X_\infty [1+3\varepsilon(X_0,Y_0;\tau)], \label{3-3-4}
\end{equation}
where
\begin{equation}
\varepsilon(X_0,Y_0;\tau) = \frac{Y_0 \exp(\mu /a^3) }{3 X_0 + Y_0} \exp(-3 X_\infty \tau),
\end{equation}
and $\mu=1/3 + a/2 + a^2 -11 a^3/16 < 0$. Making use of Eq.~(\ref{3-2-2a}) with $\gamma = 0$, we find
\begin{equation}
Y(\tau) \stackrel{\tau \to \infty}{\simeq} 3 X_\infty \varepsilon(X_0,Y_0;\tau). \label{3-3-5}
\end{equation}
Hence we have
\begin{equation}
X(\tau) \stackrel{\tau \to \infty}{\simeq} X_\infty + Y(\tau). \label{relation}
\end{equation}

\begin{figure}[ht]
\begin{center}
\includegraphics[width=0.6\columnwidth]{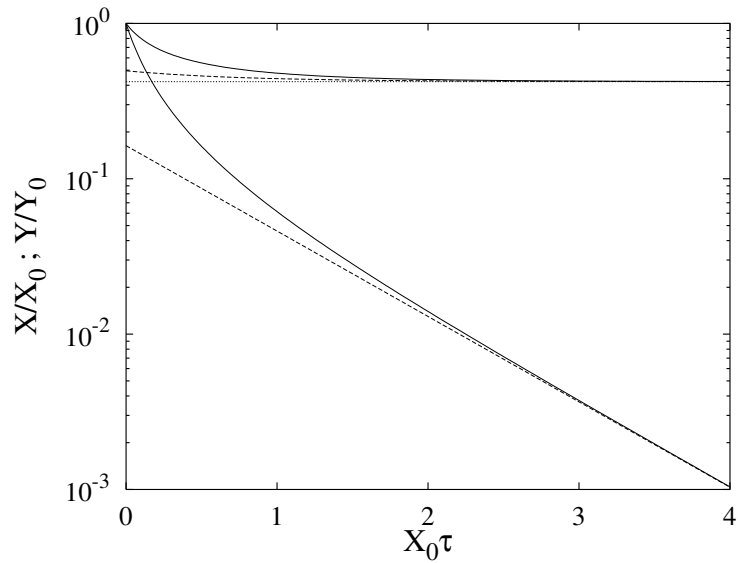}
\end{center}
\caption{\label{fig3} Linear-logarithmic plot of the numerical solution of the set 
of equations~(\ref{allsyst1}) for $X_0 = Y_0$, $\gamma = 0$ (continuous lines). 
The asymptotic relations~(\ref{3-3-4}) and~(\ref{3-3-5}) are shown by the dashed 
lines, and the asymptotic limit~(\ref{3-3-3}) by the dotted line.}
\end{figure}

There is a qualitative difference from the case $c_1 > 0$. 
As shown in Fig.~\ref{fig3}, the density of particles at rest approaches the asymptotic value $X_\infty > 0$ exponentially fast, while the density of 
moving particles goes to zero exponentially. Table~\ref{tableau1} summarizes 
the long-time behavior for the different cases.
\begin{table}[ht]
\caption{\label{tableau1}Summary of the long-time density behavior in three dimensions.}
\begin{ruledtabular}
\begin{tabular}{c|c|c|c|}
& $c_2 = c_1 > 0$ & $c_2 > c_1 \neq 0$ & $c_2 > c_1 = 0$ \\ \hline
$X(\tau)$ & $\tau^{-1}$ & $\tau^{-1}$ & $X_\infty[1+3\exp(-3 X_\infty \tau)]$ \\
$Y(\tau)$ & $\tau^{-1}$ & $\tau^{-(3+\gamma^2)/4 \gamma}$ & $X_\infty 3 \exp(-3 X_\infty \tau)$ \rule[0.2cm]{0pt}{0.0cm}
\end{tabular}
\end{ruledtabular}
\end{table}

Note that generalizing our results to any dimension $d\geq 2$ is 
straightforward (see below for the case $d=2$).  
The algebraic or exponential 
decay of the particle densities hold irrespective of $d$. 
In particular, for the general 
case $c_1>0$ the exponent of the density of ``slow'' particles is independent 
of $d$ so that $X(\tau) \stackrel{\tau \to \infty}{\sim} \tau^{-1}$. 
Finally the relation~(\ref{relation}) still holds.

%------------------------------------------------------------------------------
\section{Comparison with numerical simulations}
\label{section:md}

The analytical predictions obtained in the preceding section rely on the
validity of the molecular chaos assumption, leading to the Boltzmann
equation. It is therefore instructive to compare these predictions
to the results of molecular dynamics (MD) simulations, where 
the exact equations of motion of the particles are integrated
(see Ref.~\cite{us_three} for more details concerning the method). 

%-------------------------------------
\subsection{Analytical results in two dimensions}

MD simulations are most efficiently performed in two dimensions,
where the best statistical accuracy can be achieved. We consequently
repeat the analysis of Secs.~\ref{section2} and~\ref{section3}
for a two-dimensional system. 
Introducing the rescaled time $\tau = 2 \pi \sigma c_2 t$, one  
obtains the counterpart of Eqs. (\ref{allsyst1}) in the form
\begin{subequations}
\label{2deq}
\begin{eqnarray}
\dot{X}(\tau) &=& - 4 \gamma X(\tau)^2 - \kappa(\gamma) X(\tau) Y(\tau), \label{2deq1} \\
\dot{Y}(\tau) &=& - 4 Y(\tau)^2 - \kappa(\gamma) X(\tau) Y(\tau), \label{2deq2}
\end{eqnarray}
\end{subequations}
where $\kappa(\gamma) = \int_0^\pi d\varphi \sqrt{1+\gamma^2-2\gamma \cos\varphi}$. 
In the limit $\tau \to \infty$, the solution of the system~(\ref{2deq})  
reads
\begin{subequations}
\label{2deq0}
\begin{eqnarray}
X(\tau) &\stackrel{\tau \to \infty}{\simeq}& \tfrac{1}{4\gamma} \tau^{-1}, \label{2deq01} \\
Y(\tau) &\stackrel{\tau \to \infty}{\simeq}& \frac{V_0}{4 \gamma} 
(4 \gamma X_0)^{-1/\mu} \left( \frac{4 - \kappa}{\kappa - 4 \gamma}V_0 +1 \right)^{\nu/\mu} 
\tau^{-\kappa/4 \gamma}, \label{2deq02}
\end{eqnarray}
\end{subequations}
where $\mu = 4 \gamma/(\kappa - 4 \gamma)$ and $\nu = \kappa/(4 - \kappa)-\mu$. 
On the other hand, taking the limit $\gamma \to 0$ in Eqs.~(\ref{2deq}) 
and solving the corresponding system leads to the long-time behavior
\begin{subequations}
\label{2deqg}
\begin{eqnarray}
X(\tau) &\stackrel{\tau \to \infty}{\simeq}& X_\infty\left[1 + \varepsilon_2(X_0,Y_0;\tau) \right], \label{2deqg1} \\
Y(\tau) &\stackrel{\tau \to \infty}{\simeq}& X_\infty \, \varepsilon_2(X_0,Y_0;\tau), \label{2deqg2}
\end{eqnarray}
\end{subequations}
where 
\begin{equation}
\frac{X_\infty}{X_0}= \frac{1}{(1+V_0 \chi)^{1/\chi}} \label{xinfini}
\end{equation}
and $\chi = 4/\pi -1$. In Eqs.~(\ref{2deqg}), we have
\begin{equation}
\varepsilon_2(X_0,Y_0;\tau) = \pi V_0^{(1+V_0\chi)^{-1/\chi-1}} \exp \left( - J X_\infty/X_0 \right) \exp \left( - \pi X_\infty \tau\right)
\end{equation}
and
\begin{equation}
J = \int_0^{V_0} du \, \ln(u)  \left[ - \frac{d^2}{du^2} \left( \frac{1/\chi + V_0}{1/\chi + u} \right)^{1/\chi} \right].
\end{equation}

%-------------------------------------
\subsection{Molecular dynamics simulations}
MD simulations have been implemented with systems of typically $N=10^5$ 
to $4\times 10^5$ spheres
in two dimensions (discs). Periodic boundary conditions were enforced, and 
low densities considered, in order to minimize the excluded volume effects
discarded at the Boltzmann level (note that these effects are necessarily
transient since the density decreases with time). 

\begin{figure}[ht]
\begin{center}
\includegraphics[width=0.6\columnwidth]{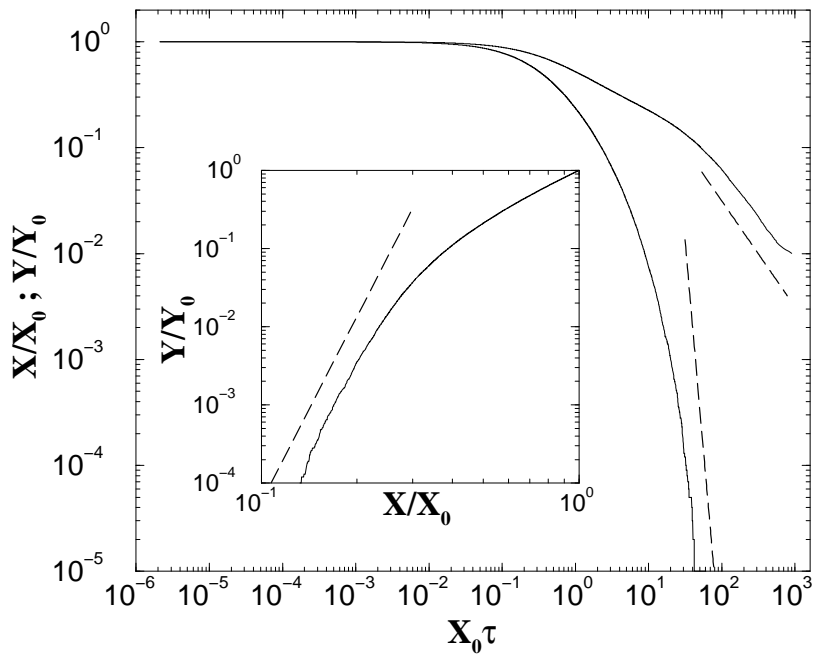}
\end{center}
\caption{Log-log plot of the densities $X$ (upper curve) and $Y$ (lower curve) as a function
of rescaled time, as obtained in the MD simulations of a two-dimensional system
with $\gamma=0.1$. The initial condition corresponds to an equimolar mixture
($X_0=Y_0$) of $N=2 \times 10^5$ particles, 
with reduced density $(X_0+Y_0)\sigma^2=0.1$ at $\tau=0$ 
(both species have the same diameter). The dashed lines have slopes $-1$ and $-7.9$
[as predicted by Eqs.~(\ref{2deq0})]. Inset: log-log plot of $Y$ as a function
of $X$, where the broken line has slope $-7.9$.  }
\label{fig4}
\end{figure}

Figure~\ref{fig4} compares the MD results obtained with $\gamma=1/10$ to
the predictions of Eqs.~(\ref{2deq0}) (for $\gamma=1/10$, the time
decay of the ``fast'' particles is governed by the exponent 
$\kappa/4\gamma \simeq 7.9$). Although the large-time behaviors for
$X$ and $Y$ are compatible with those given by Eqs.~(\ref{2deq0}), it may be
observed that the corresponding asymptotic regime is difficult to probe,
even for large systems. The parametric plot [or ``trajectory'' $Y(X)$]
shown in the inset is however in agreement with the relation
$Y \propto X^{\kappa/4\gamma}$ deduced from Eq.~(\ref{2deq0}).

\begin{figure}[ht]
\begin{center}
\includegraphics[width=0.6\columnwidth]{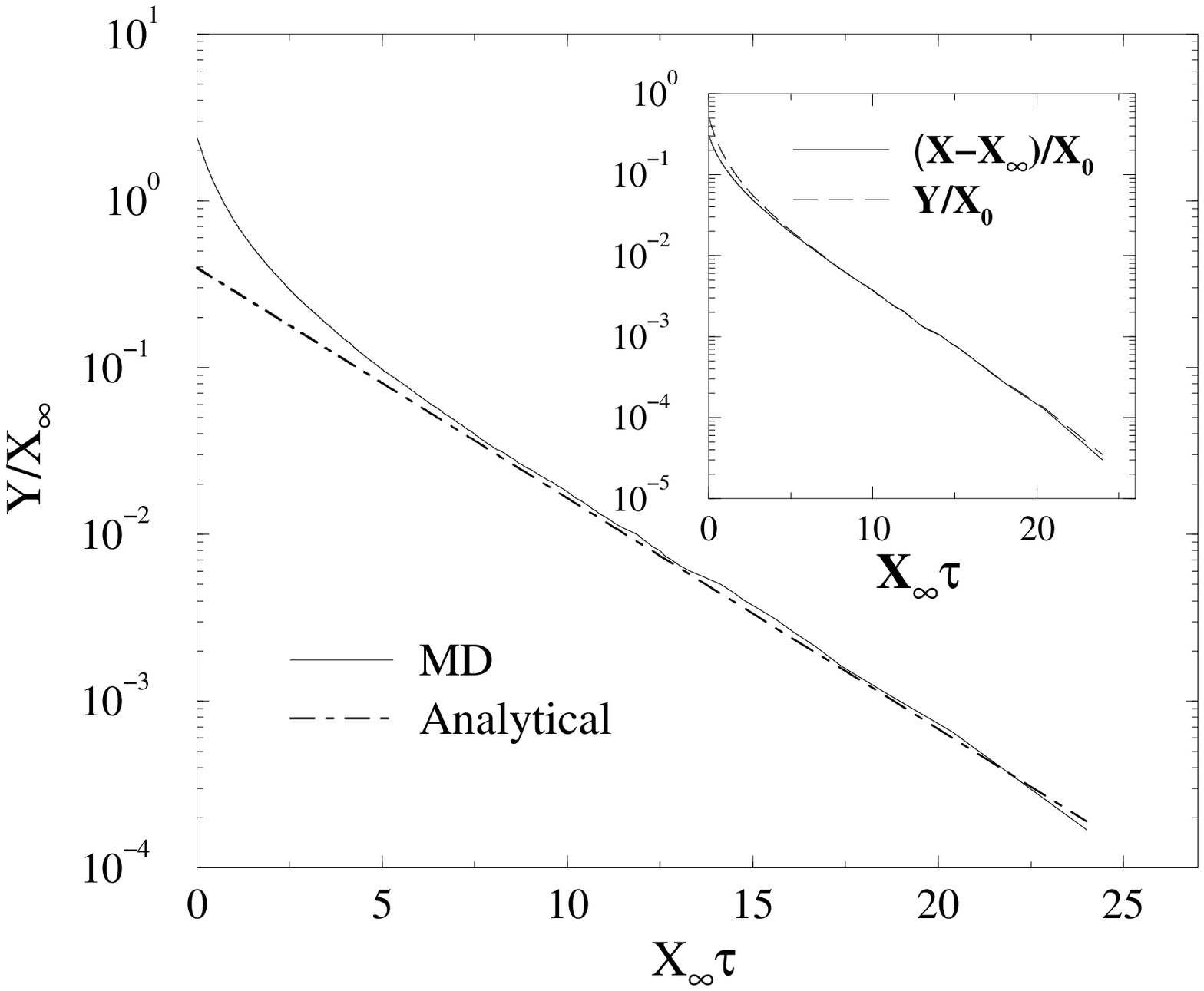}
\end{center}
\caption{Linear-logarithmic plot of the density of moving particles. Here,
$\gamma=0$, $X_0=Y_0=5\times 10^{-3}/\sigma^2$ [corresponding to a very low 
total initial packing fraction $\eta_0 \equiv \pi (X_0+Y_0)\sigma^2/4= 0.0078$]. 
The initial number of particles is $N=4\times 10^5$.
The results of MD simulations (continuous curve) 
are compared to the predictions of Eqs.~(\ref{2deqg}),
shown by the broken line. 
The inset shows that $X-X_\infty$ and $Y$ (obtained in MD) 
have asymptotically the same time decay [see  Eqs.~(\ref{2deqg})].
}
\label{fig5}
\end{figure}

We have also performed MD simulations for a mixture of
moving and motionless particles ($\gamma=0$), where it is
expected that the density $X$ of particles at rest decreases
down to a nonvanishing value $X_\infty$. 
In the situation of an
equimolar mixture ($X_0 = Y_0$), we have $V_0 = 1$ so that according to
Eq.~(\ref{xinfini}), $X_\infty/X_0\simeq 0.414$. The MD simulations
are in agreement with this scenario, and we find 
$X_\infty/X_0\simeq 0.408$ irrespective of the initial conditions for a
system with initially $N=2\times 10^5$ particles. 
The results for the time dependence of $X$ and $Y$ are displayed
in Fig.~\ref{fig5}. We conclude that the numerical simulations are again in agreement with the prediction of Boltzmann's kinetic theory.

%------------------------------------------------------------------------------
\section{Conclusions}\label{section4}
%------------------------------------------------------------------------------

We have shown that for some simple spatially homogeneous systems, 
characterized by a velocity distribution with a discrete velocity modulus spectrum, 
it is possible to find the exact solution for the nonlinear integral equation 
describing the dynamics of ballistic annihilation. These results, obtained at the level of a Boltzmann equation, have been validated by explicit comparison with molecular dynamics simulations in two dimensions.

For a single-velocity modulus distribution, the particle density of the 
model decays asymptotically as $n(t) \sim t^{-1}$, irrespective of space dimension. 
It was however rigorously shown that in one dimension, the decay is 
slower, $n(t) \sim t^{-1/2}$. 
This difference is a consequence of the fact that in one dimension strong dynamical 
correlations are created~\cite{us_one}, which invalidate the approximation underlying
Boltzmann's dynamics. In higher dimensions, the Boltzmann equation becomes
exact in the long-time limit. 

In the case of a distribution with two different finite nonzero velocity moduli, 
we found that both particle densities decay for a large time according to a power law. 
The interesting feature is that the density of the slow particles decays as $t^{-1}$, 
while the density of the fast particles decays more rapidly (e.g., as
$t^{-(3+\gamma^2)/4\gamma}$ in three dimensions), 
with a nonuniversal exponent depending continuously 
on the velocity modulus ratio $\gamma=c_1/c_2$. A rough criterion for the crossover 
time separating the short- and long-time regimes has been given in Eq.~(\ref{caractime}).

Finally, the case $c_1=0$ leads to a particularly interesting behavior. 
Independently of the initial conditions, the densities of the moving and the 
motionless particles both decrease exponentially fast; however down to a nonzero value for particles at rest. This behavior is quite different from that observed in the one-dimensional case where the initial value of the density 
of motionless particles plays an important role in the long-time regime. This 
difference between one dimension and higher dimensions reflects once again the important 
role played by the dynamically created correlations for $d=1$.

The case with motionless particles can be viewed as a problem of ballistic 
annihilation of particles with one-velocity moduli moving in a random medium 
containing immobile traps (the motionless particles) that can capture a moving 
particle and then disappear. Here again, the situation can be compared to similar 
problems in diffusion limited annihilation where the presence of traps can modify 
the long-time dynamics from a power law to an exponential decay~\cite{OBCM}.

It would be interesting to compare the above theoretical predictions with some 
experimental data. Besides growth and coarsening problems, ballistic annihilation 
could model other physical systems such as, for example, the fluorescence of laser excited 
gas atoms with quenching on contact~\cite{PRM}. However, the correspondence between 
such experimental situations and our model is not yet close enough to allow comparison. We would be highly interested in the knowledge of other physical systems that could be described by the models studied here.

\begin{acknowledgments} 
This work was partially supported by the Swiss National Science Foundation. M.D. acknowledges the support of the Swiss National Science Foundation and of the CNRS. J.P. acknowledges the hospitality at the Department of Physics of the University of Geneva. We thank P.~L.~Krapivsky for bringing our attention to Ref.~\cite{krap}.

\end{acknowledgments}
%------------------------------------------------------------------------------

\end{document}